\documentclass[showpacs]{revtex4}
\usepackage{graphicx}
\usepackage{amsmath,amssymb}
\begin{document}
\title{Motion of rotatory molecular motor and chemical reaction rate}
\author{Hiroshi Miki, Masatoshi Sato, and Mahito Kohmoto \\
\it The Institute for Solid State Physics, The University of Tokyo,\\
\it 5-1-5 Kashiwanoha, Kashiwa, Chiba 277-8581, Japan}
\date{\today}
\begin{abstract}
We examine the dependence of the physical quantities of the rotatory
molecular motor, such as the rotation velocity and the proton
translocation rate, on the chemical reaction rate using the model
based only on diffusion process.
A peculiar behavior of proton translocation is found 
and the energy transduction efficiency of the motor protein is enhanced
by this behavior. 
We give a natural explanation that this behavior is universal when
certain inequalities between chemical reaction rates hold.
That may give a clue to examine whether the motion of the molecular
motor is dominated by diffusion process or not.
\end{abstract}


\maketitle

\section{Introduction}

\ Many kinds of molecular motors are known to play essential roles of
life. The word 'molecular motor' is, in a wide sense, used to describe
protein molecules that transduce chemical energy from some source into
mechanical work (and in some case, vice versa). Molecular motors are
classified into two kinds: linear and rotatory motors. As the former,
myosins, kinesins, and dyneins, which slide along the specific filaments
in the specific direction, respectively, are known to work for 
muscle construction, transport of materials in cell, {\it etc}. by
hydrolysing ATP\cite{How,VM}. As the latter, two motors are known to date.
 The one is bacteria flagellar motor, which works as a propeller by 
gaining chemical energy from the proton gradient across the membrane.  
The other is ATP synthase which is found in membranes of living things.
It works wholly as a transducer between the energy of
proton gradient across the membrane and that of ATP synthesis from
 ADP and inorganic phosphate. It can both synthesize ATP and pump proton
 against the proton gradient by hydrolysing ATP (see
 Fig.16-28 in Ref.\cite{MCB}).  
 It is composed of two parts, the $F_o$- and the $F_1$-part, 
each is known to work as a rotatory motor
 \cite{CD}-\cite{WO}. 
 The $F_o$-part is embedded in the membrane. It contains a
 proton channel and transduces the energy of transmembrane proton
 gradient into rotatory torque when working as synthesizer. The $F_1$-
part plays a role of ATP synthesis/hydrolysis, transducing rotatory
 torque into the energy required for ATP synthesis and vice versa.   

It is believed generally that the energy transduction efficiency of
molecular motor, that is, the ratio of the energy output to input
from the energy source, is extremely high. For example, it is reported
in Ref.\cite{NYK} that the efficiency is almost 100 percent for the
$F_1$-part of ATP synthase. Thus, understanding
 the mechanism of molecular motors, especially that of
energy transduction is important not only because of their many roles of
life, but also because of the possibility that, if this mechanism is
novel, it may be applied to a new technology as a machine with high
energy transduction efficiency, even if it can be used only in a
constrained scale. In the scale of motor molecule($\sim$10-100nm), 
water molecules collides motor molecule so many times. These collisions
 make the diffusive motion of the motor molecule and thus it is natural
 to infer that this collision effect contributes to the motion of motor,
 but there is no crucial evidence that it is essential.  

Here we take up a simple model inspired by the $F_o$-part of ATP
synthase\cite{EWO,JLE} and examine the dependence of the rotation
velocity and the proton translocation rate of this model on the chemical
reaction rates.  
The efficiency of energy transduction is calculated from these
quantities.
As far as we know, there is no computation of this kind.

\section{Model}

\begin{figure}
\begin{center}
\includegraphics[width=7cm]{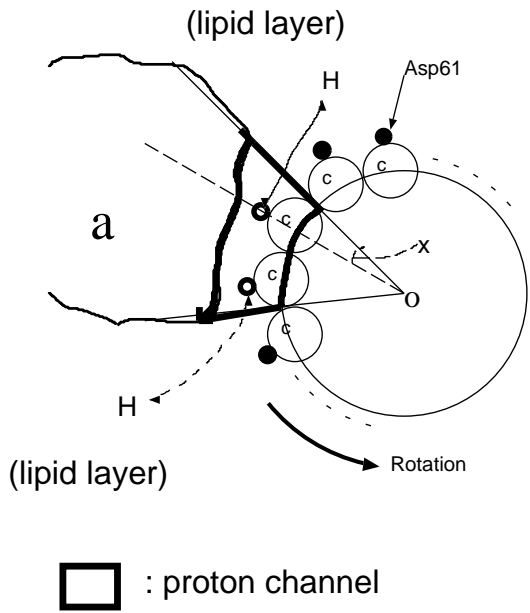}
\caption{The proton channel viewed from the basic side,
 the inner side of the membrane.
 There are the 
{\bf c}-subunits arranged in a ring and the ring rotates counterclockwise 
during synthesis. The position variable $x$ is defined as a rotation 
angle of the ring.\label{over}}
\end{center}
\end{figure}

\begin{figure}
\begin{center}
\includegraphics[width=7cm]{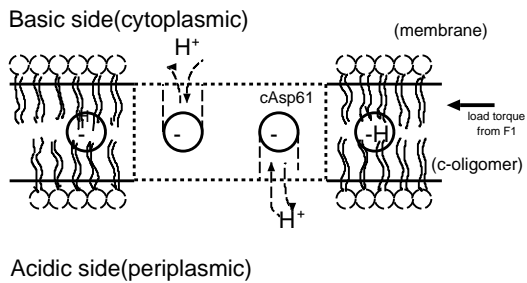}
\caption{The proton channel viewed from the {\bf a}-subunit. There are
 two proton binding sites in the channel and only these two can be
either protonated or unprotonated. The left path of
 protons is from the basic 
side and the right path is from the acidic side\cite{VA}. These proton binding 
sites move rightward during synthesis despite the leftward load from 
the $F_1$-part.\label{front}}
\end{center}
\end{figure}

 The $F_o$-part is schematically drawn in Figs.\ref{over} and \ref{front}.
It contains three types of subunits, {\bf a}, {\bf b}, and {\bf c}. As
shown in Fig.\ref{over}, the {\bf a}-subunit is fixed in the membrane. 
The {\bf b}-subunit connects the $F_o$- and $F_1$-part but is not shown
in Figs.\ref{over} and \ref{front}(see Ref.\cite{MCB}). 
The {\bf c}-subunits are arranged in a ring and this ring rotates
accompanying the proton translocation. 
There is a proton binding site in the almost middle of each {\bf c}-
subunit, Asp61, a carboxyrate. 
 A proton passes through the interface between the {\bf a}- and {\bf c}-
 subunits (but as described later, not directly), called proton channel.  
 In the proton channel,
 there are two paths to these binding sites, the left one is from the
 basic side, the inner side of the membrane, and the right is from the acidic
 side, the outer(see Fig.\ref{front})\cite{VA}.
 Proton concentration in the acidic side are kept higher than that of the
 basic side by respiratory chains, which pumps protons in the basic side
 out of the membrane.
 Roughly speaking, a proton flows into and binds
 to the right site from acidic side, goes through the membrane with
 {\bf c}-ring rotation, and dissociates from the left site and flows
 out to the basic side. That is, a proton
 passes through the proton channel not directly but via the membrane
 accompanying the rotation of the {\bf c}-ring. 
     

We investigate the above situation using the '(simply) biased diffusion
model' \cite{EWO,JLE}. 
It was first presented to explain the rotation of bacteria flagellar
motor which has similar structure to the $F_o$-part of the ATP synthase
but is one order larger in linear dimension.
The mechanism is as follows: There are $N$ proton binding sites
in the rotor (i.e. the ring is composed of $N$ {\bf c}-subunits and
$N$=9-12 in {\it {Escherichia coli}}) at same intervals and two of them
are in the proton channel. 
Only two proton binding sites are allowed to be in the
channel and each of them can be protonated or unprotonated.
The others can be only protonated due to the hydrophobicity of the membrane. 
Thus the states of the motor are determined by the protonation of the
two proton binding sites in the channel: E is the empty state where both
sites are unprotonated, R is the right state where the right site is
protonated and the left is unprotonated, L is the left state where the
left site is protonated and the right is unprotonated, and F is the full
state where both sites are protonated.
We assume that a proton does not hop between two proton binding sites.
When one of the proton binding sites in the channel comes 
to the boundary, it can move into the membrane only if it is protonated
and otherwise cannot due to hydrophobicity of the membrane. 
If there is difference from detailed balance between the rates of one
site at which a proton binds to or dissociates from it and those of the
other due to the transmembrane proton gradient, it become possible to
take out a net motion.

Dynamics of each state is described by the Langevin equation,  
\begin{equation} 
 \frac{dx}{dt}=-\gamma_i\frac{d\phi_i(x)}{dx}+R_i(t), \label{laneq}
\end{equation}
where $i$ refers to the state, $\gamma_i$ is the friction constant, $x$
is the position of the motor (see Fig.1) which is explicitly defined later.
The potential $\phi_i(x)$ represents the load torque $\tau$ from the
$F_1$ part, so $\phi_i(x)=\tau x$ \cite{NYK}.
$R_i(t)$ is the random force of Gaussian noise, which satisfies 
\begin{equation}
\langle R_i(t) \rangle=0,
\end{equation}
\begin{equation}
\langle R_i(t)R_j(t') \rangle =2D_{i}\delta_{ij}\delta(t-t'),
\end{equation} 
where $\langle \cdots \rangle$ is the time average and $D_i$ is the
diffusion constant satisfying the Einstein's relation, $D_i=\gamma_i
k_{\rm B}T$ ($T$: temperature, $k_{\rm B}$: the Boltzmann constant).
The constants $D_i$ and $\gamma_i$ depend on the states since the
different conformations have different surface areas, densities, {\it etc}.   
But for simplicity, it is assumed below that all the states have
effectively the same values, $\gamma_i=\gamma$, and $D_i=D$, since those
differences are small. 
In addition to the above single-state motions,transitions between
 different states occur.   
Thus it is convenient to adopt the formulation of the Fokker-Planck
equation equivalent to Eq.(\ref{laneq}),  
\begin{equation}
 \frac{\partial}{\partial t}{\bf{p}}(x,t)=-\frac{\partial}{\partial
 x}{\bf{\Pi}}(x,t)+{\hat{\bf K}}(x)\cdot{\bf{p}}(x,t) \label{fpeeq},
\end{equation}
\begin{equation}
 {\bf {\Pi}}(x,t)=\gamma{\hat{\bf f}}(x){\bf{p}}(x,t)-D\frac{\partial}
{\partial x}{\bf{p}}(x,t) \label{flow},
\end{equation}
where ${\hat{\bf f}}$ is the external force ${\hat{\bf f}}={\rm
diag}[\tau,\tau,\cdots,\tau]$. 
${\hat{\bf K}}$ is the transition matrix which describes changes between
states.   
The system discussed here has 4-component probability ${\bf p}(x,t)$ and
flow ${\bf \Pi}(x,t)$
\begin{eqnarray}
 {\bf{p}}(x,t)=\left[
\begin{array}{c}
p_{\rm E}(x,t)\\
p_{\rm R}(x,t)\\
p_{\rm L}(x,t) \\
p_{\rm F}(x,t)
\end{array}
\right],
\quad
 {\bf{\Pi}}(x,t)=\left[
\begin{array}{c}
\Pi_{\rm E}(x,t)\\
\Pi_{\rm R}(x,t)\\
\Pi_{\rm L}(x,t) \\
\Pi_{\rm F}(x,t)
\end{array}
\right],
\end{eqnarray}
where $p_i(x,t)$ and $\Pi_i(x,t)$($i$=E,R,L,F) describe the probability
and its flow that the motor in state $i$ is at position $x$ and at time $t$.
 The coordinate $x$ is placed as follows: The origin $O$ is set
at the center of the {\bf c}-ring and the position variable $x$ is defined
as a rotation angle (Fig.\ref{over}). The proton channel extends from
$x=0$ to $\delta$ ($\delta=2\pi/N$ and $N=12$ fixed hereafter). The
position $x=0$ is defined that the left proton binding site is at the
 left boundary of
the channel, and  $x=\delta$ is that the right site is at the right
boundary. When one site moves into the membrane, another site appears
into the channel from the other side of the membrane. In this sense, the
system has periodicity. 
These constraints of periodicity and hydrophobicity mentioned 
 above are expressed by
 giving Eq.(\ref{fpeeq}) the boundary conditions,
\begin{equation}
\Pi_{\rm E}(0,t)=\Pi_{\rm E}(\delta,t)=0,\label{nogoE}
\end{equation}
\begin{equation}
 \Pi_{\rm R}(0,t)=\Pi_{\rm L}(\delta,t)=0,\label{nogoRL}
\end{equation} 
\begin{equation}
\Pi_{\rm L}(0,t)=\Pi_{\rm R}(\delta,t), \label{peRL}
\end{equation}
and,
\begin{equation}
 \Pi_{\rm F}(0,t)=\Pi_{\rm F}(\delta,t). \label{peF}
\end{equation}

The transition rate matrix is written as
 \begin{equation}
  \hat{\bf K}=\left[
\begin{array}{cccc}
-(k^{\rm R}_{\rm in}+k^{\rm L}_{\rm in})&k^{\rm R}_{\rm out}
&k^{\rm L}_{\rm out}&0\\
k^{\rm R}_{\rm in}&-(k^{\rm R}_{\rm out}+k^{\rm L}_{\rm in})
&0&k^{\rm L}_{\rm out}\\
k^{\rm L}_{\rm in}&0&-(k^{\rm R}_{\rm in}+k^{\rm L}_{\rm in})
&k^{\rm R}_{\rm out}\\
0&k^{\rm L}_{\rm in}&k^{\rm R}_{\rm in}
&-(k^{\rm R}_{\rm out}+k^{\rm L}_{\rm out})\\
\end{array}
\right].
\end{equation}
There are a few assumptions in the above expression of the transition
rate matrix : 1) The chemical reaction, that is, binding and/or
dissociation of proton to/from the sites, is sufficiently fast compared
with the motion of the motor protein. 
This is justified because the mass and size of a proton is much smaller
than those of the motor so the diffusion coefficient of
a proton is much larger than that of the motor protein. 
2) There is no
correlation between the reaction of the left site and that of the right
site. 
3) The reaction rates are independent of the position of the
motor $x$. This assumption may have to be modified to take account of
other influences, for instance, interaction between residues\cite{EWO}.
 But we neglect the influences like these for simplicity.
 In this case, for example, it is thought that $k_{\rm ER}$, the rate at which
 state E switches to state R and $k_{\rm LF}$, the one at which L switches to 
 F are both written by $k_{\rm in}^{\rm R}$, the one at which a proton
 binds to the empty right site. So similarly we set,
\begin{eqnarray}
 k_{\rm ER}=k_{\rm LF}=k^{\rm R}_{\rm in},
\\
 k_{\rm EL}=k_{\rm RF}=k^{\rm L}_{\rm in},
\\
 k_{\rm RE}=k_{\rm FL}=k^{\rm R}_{\rm out},
\end{eqnarray}
and,
\begin{eqnarray}
 k_{\rm LE}=k_{\rm FR}=k^{\rm L}_{\rm out}.
\end{eqnarray}
For the motor to work, the detailed balance should be violated due to the 
 free energy acquired from a proton passage of the channel, $\Delta G$.
Thus the ratio of transition rate from L to R via E or F to that from R to L
is written as, 
\begin{eqnarray}
\frac{({\rm L}\rightarrow {\rm R})}{({\rm R}\rightarrow {\rm L})}=
\frac{k_{\rm LE}k_{\rm ER}+k_{\rm LF}k_{\rm FR}}
{k_{\rm RE}k_{\rm EL}+k_{\rm RF}k_{\rm FL}}=
\frac{k^{\rm R}_{\rm in}k^{\rm L}_{\rm out}}
{k^{\rm L}_{\rm in}k^{\rm R}_{\rm out}}=\exp[\Delta G/k_{\rm B}T],
\end{eqnarray}
where $\Delta G$ is a function of $\Delta pH$($\Delta pH=pH_{\rm B}
-pH_{\rm A}$,
A, B denotes the acidic and basic side, respectively) and $V$, the 
membrane potential accompanying this difference of proton concentration
 and written as,
\begin{eqnarray}
\Delta G =V+k_{\rm B}T\ln[10^{\Delta pH}].
\end{eqnarray}

Under these assumptions, this model can be solved
analytically. The transition rate matrix $\hat {\bf K}$ is diagonalized 
by the matrix $\hat {\bf Q}$,
 \begin{equation}
  \hat{\bf Q}=\left[
\begin{array}{cccc}
k^{\rm R}_{\rm out}k^{\rm L}_{\rm out}
&-k^{\rm R}_{\rm out}&-k^{\rm L}_{\rm out}&1\\
k^{\rm R}_{\rm in}k^{\rm L}_{\rm out}
&-k^{\rm R}_{\rm in}&k^{\rm L}_{\rm out}&-1\\
k^{\rm R}_{\rm out}k^{\rm L}_{\rm in}
&k^{\rm R}_{\rm out}&-k^{\rm L}_{\rm in}&-1\\
k^{\rm R}_{\rm in}k^{\rm L}_{\rm in}
&k^{\rm R}_{\rm in}&k^{\rm L}_{\rm in}&1\\
\end{array}
\right],
\end{equation}
 \begin{equation}
\hat{\bf Q}^{-1}=\frac{1}{(k^{\rm L}_{\rm in}+k^{\rm L}_{\rm out})
(k^{\rm R}_{\rm in}+k^{\rm R}_{\rm out})}\left[
\begin{array}{cccc}
1&1&1&1\\
-k^{\rm L}_{\rm in}&-k^{\rm L}_{\rm in}
&k^{\rm L}_{\rm out}&k^{\rm L}_{\rm out}\\
-k^{\rm R}_{\rm in}&k^{\rm R}_{\rm out}
&-k^{\rm R}_{\rm in}&k^{\rm R}_{\rm out}\\
k^{\rm L}_{\rm in}k^{\rm R}_{\rm in}&-k^{\rm L}_{\rm in}k^{\rm R}_{\rm out}
&-k^{\rm L}_{\rm out}k^{\rm R}_{\rm in}
&k^{\rm L}_{\rm out}k^{\rm R}_{\rm out}\\
\end{array}
\right],
\end{equation}
and
 \begin{equation}
  \hat{\bf Q}^{-1}\hat{\bf K}\hat{\bf Q}=\left[
\begin{array}{cccc}
0&&&\\
&-(k^{\rm L}_{\rm in}+k^{\rm L}_{\rm out})&&\\
&&-(k^{\rm R}_{\rm in}+k^{\rm R}_{\rm out})&\\
&&&-(k^{\rm R}_{\rm in}+k^{\rm L}_{\rm in}
+k^{\rm R}_{\rm out}+k^{\rm L}_{\rm out})\\
\end{array}
\right].
\end{equation}
Then in a steady state, Eq.(\ref{fpeeq}) is reduced to
\begin{flushleft}
\[
 0=-\frac{d}{dx}\left[\gamma\tau-D\frac{d}{dx}\right](\hat{\bf
 Q}^{-1}{\bf p}(x))
\]
\end{flushleft}
\begin{equation}
+\left[
\begin{array}{cccc}
0&&&\\
&-(k^{\rm L}_{\rm in}+k^{\rm L}_{\rm out})&&\\
&&-(k^{\rm R}_{\rm in}+k^{\rm R}_{\rm out})&\\
&&&-(k^{\rm R}_{\rm in}+k^{\rm L}_{\rm in}
+k^{\rm R}_{\rm out}+k^{\rm L}_{\rm out})\\
\end{array}
\right](\hat{\bf Q}^{-1}{\bf p}(x)).\label{reduced}
\end{equation}
The solution of Eq.(\ref{reduced}) is,
\begin{equation}
(\hat{\bf Q}^{-1}{\bf p}(x)) =\left[
\begin{array}{c}
C_1+C_2e^{\xi x}\\
C_3e^{\eta^+_{\rm L}x}+C_4e^{\eta^-_{\rm L}x}\\
C_5e^{\eta^+_{\rm R}x}+C_6e^{\eta^-_{\rm R}x}\\
C_7e^{\eta^+_{\rm LR}x}+C_8e^{\eta^-_{\rm LR}x}\\
\end{array}
\right],
\end{equation}
where $C_i$ ($i=1,2,\cdots,8$) are integral constants and,
\begin{equation}
 \xi=\frac{\gamma\tau}{D}, \label{xi}
\end{equation}
\begin{equation}
 \eta_{\rm L}^{\pm}
=\frac{\gamma\tau\pm\sqrt{(\gamma\tau)^2
+4D(k^{\rm L}_{\rm in}+k^{\rm L}_{\rm out})}}{2D}, \label{etal}
\end{equation}
\begin{equation}
 \eta_{\rm R}^{\pm}
=\frac{\gamma\tau\pm\sqrt{(\gamma\tau)^2
+4D(k^{\rm R}_{\rm in}+k^{\rm R}_{\rm out})}}{2D}, \label{etar}
\end{equation}
and,
\begin{equation}
 \eta_{\rm LR}^{\pm}=
\frac{\gamma\tau\pm\sqrt{(\gamma\tau)^2
+4D(k^{\rm L}_{\rm in}+k^{\rm L}_{\rm out}
+k^{\rm R}_{\rm in}+k^{\rm R}_{\rm out})}}{2D} \label{etalr}.
\end{equation}
Therefore, we obtain
\[
{\bf p}(x)=\hat{\bf Q}(\hat{\bf Q}^{-1}{\bf p}(x))
\]
\begin{equation}
=\left[
\begin{array}{c}
k^{\rm R}_{\rm out}k^{\rm L}_{\rm out}(C_1+C_2e^{\xi x})
-k^{\rm R}_{\rm out}(C_3e^{\eta^+_{\rm L}x}+C_4e^{\eta^-_{\rm L}x})
-k^{\rm L}_{\rm out}(C_5e^{\eta^+_{\rm R}x}+C_6e^{\eta^-_{\rm R}x})
+C_7e^{\eta^+_{\rm LR}x}+C_8e^{\eta^-_{\rm LR} x}\\
k^{\rm R}_{\rm in}k^{\rm L}_{\rm out}(C_1+C_2e^{\xi x})
-k^{\rm R}_{\rm in}(C_3e^{\eta^+_{\rm L}x}+C_4e^{\eta^-_{\rm L}x})
+k^{\rm L}_{\rm out}(C_5e^{\eta^+_{\rm R}x}+C_6e^{\eta^-_{\rm R}x})
-C_7e^{\eta^+_{\rm LR} x}-C_8e^{\eta^-_{\rm LR} x}\\
k^{\rm R}_{\rm out}k^{\rm L}_{\rm in}(C_1+C_2e^{\xi x})
+k^{\rm R}_{\rm out}(C_3e^{\eta^+_{\rm L}x}+C_4e^{\eta^-_{\rm L}x})
-k^{\rm L}_{\rm in}(C_5e^{\eta^+_{\rm R}x}+C_6e^{\eta^-_{\rm R}x})
-C_7e^{\eta^+_{\rm LR} x}-C_8e^{\eta^-_{\rm LR} x}\\
k^{\rm R}_{\rm in}k^{\rm L}_{\rm in}(C_1+C_2e^{\xi x})
+k^{\rm R}_{\rm in}(C_3e^{\eta^+_{\rm L}x}+C_4e^{\eta^-_{\rm L}x})
+k^{\rm L}_{\rm in}(C_5e^{\eta^+_{\rm R}x}+C_6e^{\eta^-_{\rm R}x})
+C_7e^{\eta^+_{\rm LR}x}+C_8e^{\eta^-_{\rm LR}x}\\
\end{array}
\right], 
\end{equation}
and,
\begin{eqnarray}
 {\bf \Pi}(x)&=&\hat{\bf Q}\left[\gamma\tau-D\frac{d}{dx}\right]
(\hat{\bf Q}^{-1}{\bf p}(x))
\nonumber\\
 &=&\hat{\bf Q}\left[
\begin{array}{c}
C_1\gamma\tau \\
C_3D\eta^-_{\rm L}e^{\eta^+_{\rm L}x}
+C_4D\eta^+_{\rm L}e^{\eta^-_{\rm L}x}\\
C_5D\eta^-_{\rm R}e^{\eta^+_{\rm R}x}
+C_6D\eta^+_{\rm R}e^{\eta^-_{\rm R}x}\\
C_7D\eta^-_{\rm LR}e^{\eta^+_{\rm LR}x}
+C_8D\eta^-_{\rm LR}e^{\eta^-_{\rm LR}x}\\
\end{array}
\right].
\label{Pisol}
\end{eqnarray}
The integral constants $C_i$'s are determined to satisfy
 the boundary conditions (\ref{nogoE})-(\ref{peF}).
 Five of these six boundary conditions are
independent but the rest one is not because from Eq.(\ref{fpeeq}),
the following relation is always satisfied in the steady state,
\begin{eqnarray}
0&=&
 \frac{\partial}{\partial t}\int^{\delta}_0 dx \sum_i p_i(x,t)
\nonumber\\
 &=&-\int^{\delta}_0 dx \sum_i \frac{\partial}{\partial x}\Pi_i(x,t)
\nonumber\\
 &=&\sum_i (\Pi_i(0,t)-\Pi_i(\delta,t)).
\end{eqnarray}
In addition to these boundary conditions, the
periodicity of the probability at the boundaries,
\begin{equation}
p_{\rm R}(\delta)=p_{\rm L}(0),\label{pRL}
\end{equation}
\begin{equation}
p_{\rm F}(\delta)=p_{\rm F}(0),\label{pF}
\end{equation}
and the normalization condition
\begin{equation}
\int^{\delta}_0 dx \sum_i p_i(x)=1 \label{normal}
\end{equation}
are necessary in order to close the equations for $C_i$'s.
 Using these conditions $C_i$'s are determined as,
\begin{eqnarray}
 {\bf c} ={\bf{\hat M}}^{-1} {\bf s},
\end{eqnarray}
where
\begin{eqnarray}
 \bf c=\left[
\begin{array}{c}
C_1\\
C_2\\
\vdots\\
C_8\\
\end{array}
\right],
\end{eqnarray}
\begin{eqnarray}
 {\bf s}=\frac{1}{(k_{\rm in}^{\rm L}+k_{\rm out}^{\rm L})
(k_{\rm in}^{\rm R}+k_{\rm out}^{\rm R})}\left[
\begin{array}{c}
0\\
0\\
\vdots\\
1\\
\end{array}
\right],
\end{eqnarray}
and,
\begin{eqnarray}
&&\hspace{20ex} \bf{\hat M}=
\nonumber\\
&& \left[
\begin{array}{cccc}
k_{\rm in}^{\rm R}k_{\rm out}^{\rm L}\gamma\tau&0
&-k_{\rm in}^{\rm R}D\eta_{\rm L}^-&-k_{\rm in}^{\rm R}D\eta_{\rm L}^+\\
k_{\rm out}^{\rm R}k_{\rm in}^L\gamma\tau&0
&k_{\rm out}^{\rm R}D\eta_{\rm L}^-e^{\eta_{\rm }L^+\delta}
&k_{\rm out}^{\rm R}D\eta_{\rm L}^+e^{\eta_{\rm L}^-\delta}\\
k_{\rm out}^{\rm R}k_{\rm out}^{\rm L}\gamma\tau&0
&-k_{\rm out}^{\rm R}D\eta_{\rm L}^-&-k_{\rm out}^{\rm R}D\eta_{\rm L}^+\\
k_{\rm out}^{\rm R}k_{\rm out}^{\rm L}\gamma\tau&0
&-k_{\rm out}^{\rm R}D\eta_{\rm L}^-e^{\eta_{\rm L}^+\delta}&
-k_{\rm out}^{\rm R}D\eta_{\rm L}^+e^{\eta_{\rm L}^-\delta}\\
k_{\rm in}^{\rm R}k_{\rm out}^{\rm L}-k_{\rm out}^{\rm R}k_{\rm in}^{\rm L}
&k_{\rm in}^{\rm R}k_{\rm out}^{\rm L}e^{\xi\delta}-k_{\rm out}^{\rm R}
k_{\rm in}^{\rm L}
&-k_{\rm in}^{\rm R}e^{\eta_{\rm L}^+\delta}-k_{\rm out}^{\rm R}
&-k_{\rm in}^{\rm R}e^{\eta_{\rm L}^-\delta}-k_{\rm out}^{\rm R}\\
0&k_{\rm in}^{\rm R}k_{\rm in}^{\rm L}(1-e^{\xi\delta})
&k_{\rm in}^{\rm R}(1-e^{\eta_{\rm L}^+\delta})&
k_{\rm in}^R(1-e^{\eta_{\rm L}^-\delta})\\
0&\xi k_{\rm in}^{\rm R}k_{\rm in}^{\rm L}(1-e^{\xi\delta})
&k_{\rm in}^{\rm R}\eta_{\rm L}^+(1-e^{\eta_{\rm L}^+\delta})&
k_{\rm in}^{\rm R}\eta_{\rm L}^-(1-e^{\eta_{\rm L}^-\delta})\\
\delta&(e^{\xi\delta}-1)/\xi&0&0\\
\end{array}
\right.
\nonumber\\
&&\left.
\begin{array}{cccc}
k_{\rm out}^{\rm L}D\eta_{\rm R}^-&k_{\rm out}^{\rm L}D\eta_{\rm R}^+
&-D\eta_{\rm LR}^-&-D\eta_{\rm LR}^+\\
-k_{\rm in}^{\rm L}D\eta_{\rm R}^-e^{\eta_{\rm R}^+\delta}
&-k_{\rm in}^{\rm L}D\eta_{\rm R}^+e^{\eta_{\rm R}^-\delta}
&-D\eta_{\rm LR}^-e^{\eta_{\rm LR}^+\delta}
&-D\eta_{\rm LR}^+e^{\eta_{\rm LR}^-\delta}\\
-k_{\rm out}^{\rm L}D\eta_{\rm R}^-&-k_{\rm out}^{\rm L}D\eta_{\rm R}^+
&D\eta_{\rm LR}^-&D\eta_{\rm LR}^+\\
-k_{\rm out}^{\rm L}D\eta_{\rm R}^-e^{\eta_{\rm R}^+\delta}
&-k_{\rm out}^{\rm L}D\eta_{\rm R}^+e^{\eta_{\rm R}^-\delta}
&D\eta_{\rm LR}^-e^{\eta_{\rm LR}^+\delta}
&D\eta_{\rm LR}^+e^{\eta_{\rm LR}^-\delta}\\
k_{\rm out}^{\rm L}e^{\eta_{\rm R}^+\delta}+k_{\rm in}^{\rm L}
&k_{\rm out}^{\rm L}e^{\eta_{\rm R}^-\delta}+k_{\rm in}^{\rm L}
&1-e^{\eta_{\rm LR}^+\delta}
&1-e^{\eta_{\rm LR}^-\delta}\\
k_{\rm in}^{\rm L}(1-e^{\eta_{\rm R}^+\delta})
&k_{\rm in}^{\rm L}(1-e^{\eta_{\rm R}^-\delta})
&1-e^{\eta_{\rm LR}^+\delta}
&1-e^{\eta_{\rm LR}^-\delta}\\
k_{\rm in}^{\rm L}\eta_{\rm R}^+(1-e^{\eta_{\rm R}^+\delta})
&k_{\rm in}^{\rm L}\eta_{\rm R}^-(1-e^{\eta_{\rm R}^-\delta})
&\eta_{\rm LR}^+(1-e^{\eta_{\rm LR}^+\delta})
&\eta_{\rm LR}^-(1-e^{\eta_{\rm LR}^-\delta})\\
0&0&0&0\\
\end{array}
\right].
\end{eqnarray}

We find useful formulas for the probability and the flow in state $i$. 
For the probalities, the following relations hold in the steady state,
\begin{eqnarray}
0&=&-(k^{\rm R}_{\rm in}+k^{\rm L}_{\rm in}){\bar p_{\rm E}}
+k^{\rm R}_{\rm out}{\bar p_{\rm R}}
+k^{\rm L}_{\rm out}{\bar p_{\rm L}},  
\label{stE}\\
\Pi_{\rm R}(\delta)&=&k^{\rm R}_{\rm in}{\bar p_{\rm E}}
-(k^{\rm R}_{\rm out}+k^{\rm L}_{\rm in}){\bar p_{\rm R}}
+k^{\rm L}_{\rm out}{\bar p_{\rm F}}, 
\label{stR}\\
-\Pi_{\rm R}(\delta)&=&k^{\rm L}_{\rm in}{\bar p_{\rm E}}
-(k^{\rm R}_{\rm in}+k^{\rm L}_{\rm out}){\bar p_{\rm L}}
+k^{\rm R}_{\rm out}{\bar p_{\rm F}}, 
\label{stL}\\
0&=&k^{\rm L}_{\rm in}{\bar p_{\rm R}}
+k^{\rm R}_{\rm in}{\bar p_{\rm L}}
-(k^{\rm R}_{\rm out}+k^{\rm L}_{\rm out}){\bar p_{\rm F}},
\label{stF}
\end{eqnarray}
where
\begin{equation}
{\bar p_i}=\int^{\delta}_{0}dxp_i(x).
\end{equation}
The relations above are obtained by integrating Eq.(\ref{fpeeq}) over $x$ and using the boundary conditions (\ref{nogoE})-(\ref{peF}).
For the flow of probavility, the following relation holds,
\begin{eqnarray}
\bar {\Pi}_i 
&\equiv & \int^{\delta}_0 dx \Pi_i(x)\nonumber\\
&=&\gamma \tau \bar{p_i}(x)
+D[p_i(0)-p_i(\delta)], \label{intflow}
\end{eqnarray} 
which are obtained by integrating Eq.(\ref{flow}) with respect to $x$.

Let us now derive formulas for the observables; the average rotation
velocity $\langle v \rangle$, the proton translocation rate $N(H^{+})$,
and the efficiency of energy transduction.   
The average rotation velocity $\langle v \rangle$ is given by the total
sum of the integrated flow of probability ,
\begin{eqnarray}
\langle v \rangle=\sum_{i={\rm E,R,L,F}} \bar{\Pi}_i.
\end{eqnarray}
Using Eq.(\ref{Pisol}), we obtain
\begin{eqnarray}
\langle v \rangle=\gamma\tau\delta C_1(k^{\rm L}_{\rm in}+k^{\rm L}_{\rm out})
(k^{\rm R}_{\rm in}+k^{\rm R}_{\rm out}). \label{v}
\end{eqnarray}
The proton translocation rate $N({\rm H}^+)$ is given as follows:
A proton goes out from the left binding site to the cytoplasma when the
state L becomes E or F becomes R. And when E becomes L or R becomes F, a
proton comes back to the left binding site. So the flow of the left
binding site $J_{\rm L}$ is defined as
\begin{eqnarray}
J_{\rm L}=k^{\rm L}_{\rm out}{\bar p_{\rm L}}
-k^{\rm L}_{\rm in}{\bar p_{\rm E}}
+k^{\rm L}_{\rm out}{\bar p_{\rm F}}
-k^{\rm L}_{\rm in}{\bar p_{\rm R}}.
\end{eqnarray} 
In a similar way, the flow of the right binding site is written as
\begin{eqnarray}
J_{\rm R}=-k^{\rm R}_{\rm out}{\bar p_{\rm R}}
+k^{\rm R}_{\rm in}{\bar p_{\rm E}}
-k^{\rm R}_{\rm out}{\bar p_{\rm F}}
+k^{\rm R}_{\rm in}{\bar p_{\rm L}}.
\end{eqnarray}
For the system to be in a steady state, $J_{\rm L}$ and $J_{\rm
R}$ must be equivalent and we obtain
\begin{eqnarray}
N({\rm H}^+)
&=&J_{\rm L}=J_{\rm R}
\nonumber\\
&=&\frac{1}{2}\{J_{\rm L}+J_{\rm R}\} 
\nonumber\\
&=&\frac{1}{2}[(k^{\rm R}_{\rm in}-k^{\rm L}_{\rm in}){\bar p_{\rm E}}
-(k^{\rm R}_{\rm out}+k^{\rm L}_{\rm in}){\bar p_{\rm R}}
+(k^{\rm R}_{\rm in}+k^{\rm L}_{\rm out}){\bar p_{\rm L}}
+(k^{\rm L}_{\rm out}-k^{\rm R}_{\rm out}){\bar p_{\rm F}}].
\end{eqnarray}
Using Eqs. (\ref{stR}) and (\ref{stL}), we find that this $N({\rm H}^+)$ is
simply given by $\Pi_R(\delta)$ and evaluated as, 
\begin{eqnarray}
N({\rm H}^+)&=&\frac{1}{2}\left[{\rm Eq.}(\ref{stR})-{\rm Eq.}(\ref{stL})
\right]
\nonumber\\
&=&\Pi_{\rm R}(\delta)(=\Pi_{\rm L}(0))
\nonumber\\
&=&k^{\rm R}_{\rm in}k^{\rm L}_{\rm out} C_1 \gamma \tau
-k^{\rm R}_{\rm in}(C_3D \eta_{\rm L}^-e^{\eta_{\rm L}^+ \delta}
+C_4D \eta_{\rm L}^+e^{\eta_{\rm L}^- \delta})
\nonumber\\
&&+k^{\rm L}_{\rm out}(C_5D \eta_{\rm R}^-e^{\eta_{\rm R}^+ \delta}
+C_6D \eta_{\rm R}^+e^{\eta_{\rm R}^- \delta})
-(C_7D \eta_{\rm LR}^-e^{\eta_{\rm LR}^+ \delta}
+C_8D \eta_{\rm LR}^+e^{\eta_{\rm LR}^- \delta}). \nonumber\\
\label{nhp}
\end{eqnarray}
Finally, we define the efficiency of energy transduction as follows,
\begin{equation}
e \equiv \frac{-\tau\langle v \rangle}{\Delta G \cdot N({\rm H}^+)}.
\label{effc}
\end{equation}
This definition coincides with that given in \cite{Sekimoto}.

\section{Results}

We investigate the dependence of the quantities, $\langle v
\rangle$, $N({\rm H}^+)$ and $e$ on the transition rate $\hat{\bf K}$
using the above analytical solution. 
Parameters used for
calculation are, according to Ref.\cite{EWO}, 
$D=2\times10^4$rad${}^2$sec${}^{-1}$, $k_{\rm B}T=$4pN$\cdot$nm,
 $\gamma=5\times10^3$rad${}^2$sec${}^{-1}$pN${}^{-1}$nm${}^{-1}$ and
 $\tau = -40$pN$\cdot$nm.
The transition rates are given as
\begin{equation}
\left[
\begin{array}{c}
k^{\rm R}_{\rm in}\\
k^{\rm L}_{\rm in}\\
k^{\rm R}_{\rm out}\\
k^{\rm L}_{\rm out}
\end{array} 
\right]
=10^K
\left[
\begin{array}{c}
10^{-pH_{\rm A}}e^{\phi/k_{\rm B}T}\\
10^{-pH_{\rm B}}e^{\phi/k_{\rm B}T}\\
10^{-pK_a}e^{-V/2k_{\rm B}T}\\
10^{-pK_a}e^{V/2k_{\rm B}T}\\
\end{array}
\right],\label{chem}
\end{equation}
where $pH_{\rm A}=7.0$ and $pH_{\rm B}=8.4$ are the proton concentrations
 of each side (A:acidic, B:basic, respectively), $pK_a$ is the acidity
 of proton binding site, $\phi=2.3k_{\rm B}T$ is the surface effect, and
 $V=5.6k_{\rm B}T$ is the membrane potential (the proton binding site
 is assumed to be in the middle of the membrane).
 The overall factor $10^K$ means
 the proton absorption rate of the path. Since it depends on the proton
 diffusion coefficient and the surface area of the channel, we assume
 that this factor is common value for all transition rates.
 Variation of $K$ is also regarded as varying $pH$ of each side and $pK_a$,
 preserving the differences of each other. 

\begin{figure}
\begin{center}
\includegraphics[width=7cm]{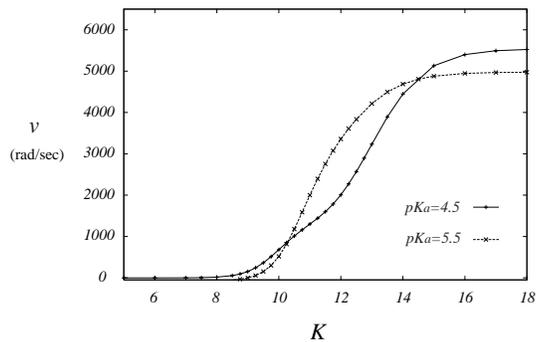}
\caption{The rotation velocity $\langle v \rangle$.
 $K$ of the abscissa is that of Eq.(\ref{chem}).
 The solid line with +'s is the plot for $pK_a$=4.5
 and the broken line with $\times$'s is for $pK_a$=5.5.
There is almost no rotation in the region of small $K$.
The faster the chemical reaction, the more this quantity 
increases monotonically and it saturates finally.  \label{rot}}
\end{center}
\end{figure}

\begin{figure}
\begin{center}
\includegraphics[width=7cm]{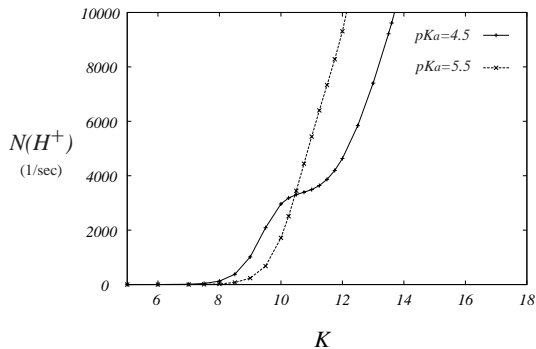}
\caption{The proton translocation rate $N({\rm H}^+)$.
 The solid line with +'s is the plot for $pK_a$=4.5
 and the broken line with $\times$'s is for $pK_a$=5.5.
There is almost no translocation observed for small $K$ in both cases.
 For $pK_a=4.5$, there exists a region where the increase becomes very slow
($K \sim$10-12). At large $K$, it diverges as $10^{K/2}$. \label{nh}}
\end{center}
\end{figure}

\begin{figure}
\begin{center}
\includegraphics[width=7cm]{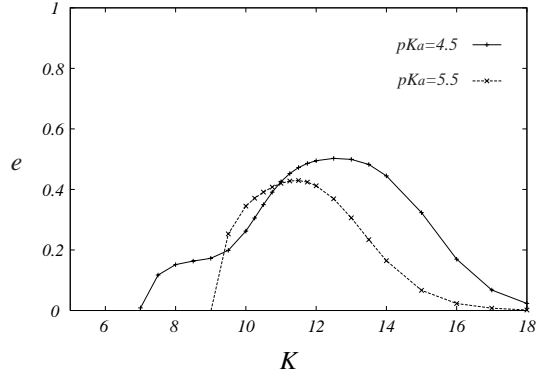}
\caption{The transduction efficiency $e$.
 The solid line with +'s is the plot for $pK_a$=4.5
 and the broken line with $\times$'s is for $pK_a$=5.5.
 It has a peak at $K \sim$13 and $e \sim$0.5 for $pK_a$=4.5,
 and $K \sim$12 and $e\sim$0.4 for $pK_a$=5.5. For large K, it
 converges to 0 reflecting the fact that $\langle v \rangle$
 converges and $N({\rm H}^+)$ diverges.  \label{eff}}
\end{center}
\end{figure}

We plot the dependence of $\langle v \rangle$, $N({\rm H}^+)$ and $e$ on
$K$ in Figs.\ref{rot}- \ref{eff}.   
For the rotation velocity, $\langle v \rangle$, no qualitative
difference dependent on $pK_a$ is found. 
There is almost no rotation in the desirable direction when the
chemical reaction rate is low (namely when $K$ is small).   
Then the velocity monotonically increases as $K$ increases and saturates
when the reaction is very fast. 
For the proton translocation rate, $N({\rm H}^+)$, a qualitative
difference dependent on $pK_a$ is found though almost no translocation
is observed during small $K$ in both cases. 
For $pK_a$=5.5, $N({\rm H}^+)$ increases monotonically.
On the other hand, for $pK_a$=4.5, while it increases monotonically
also, but there exists a region where the increase becomes very slow($K
\sim$10-12).  
In both $pK_a$'s, $N({\rm H}^+)$ diverges as $10^{K/2}$ when $K$
becomes large enough. 
The efficiency, $e$, reflects the above results of $\langle v\rangle$ 
and $N({\rm H}^+)$.  
In both $pK_a$'s, it increases as $K$ increases at first. 
Then it reaches a peak value (at $K\sim$12 for $pK_a$=5.5 and at $K\sim$13 for
$pK_a$=4.5) and decreases monotonically afterward.
The efficiency goes to zero as $K$ diverges reflecting the fact
that $\langle v \rangle$ converges to the finite value but $N({\rm
H}^+)$ diverges for large $K$.   
Note that the peak value of the efficiency for $pK_a=4.5$ is enhanced and
is about 25 percent larger than that for $pK_a=5.5$. 
This comes from the slow increase of $N({\rm H}^+)$ for $pK_a=4.5$
described above.  


Inferring from the forms of Eqs.(\ref{xi})-(\ref{etalr}),
the key to account for these results are thought to be the inequalities
between the chemical reaction rates $k^i_j$'s($i$=E,R,L,F and $j$=in,out)
and the reciprocal of the relaxation time of this convection-diffusion
system, $(\gamma\tau)^2/D$. 
The inequalities between $k_j^i$'s for $pK_a$=5.5 are different to those
for $pK_a$=4.5. For $pK_a$=5.5, the inequalities are 
$k^{\rm L}_{\rm out}>k^{\rm R}_{\rm in}
>k^{\rm R}_{\rm out}>k^{\rm L}_{\rm in}$,
while for $pK_a$=4.5,$k^{\rm L}_{\rm out}>k^{\rm R}_{\rm out}
>k^{\rm R}_{\rm in}>k^{\rm L}_{\rm in}$. 
Under the parameters used here, which are plausible in living body,
only the above two sets of inequalities are possible for the model
to work well. 
Note that the inequalities $k^{\rm R}_{\rm in}>k^{\rm L}_{\rm in}$ and
$k^{\rm L}_{\rm out}>k^{\rm R}_{\rm out}$ hold since the proton
concentration in the acidic side are higher than that in the basic side
and the difference makes the transmembrane electrostatic potential, and
also that $k^{\rm L}_{\rm out}>k^{\rm L}_{\rm in}$ is required for the
motor not to diffuse leftward by the load torque $\tau$ from the $F_1$-part.  

While the above two sets of inequalities between $k^i_j$'s do not depend
 on $K$, but the inequalities between $(\gamma \tau)^2/D$ and $k^i_j$'s do
 depend on $K$. And $k^i_j$'s which are larger than $(\gamma \tau)^2/D$
 are relevant to the dynamics of the system. When $K$ is small enough, 
 $(\gamma \tau)^2/D$ is the largest among them. Thus all the transition
 rates between the states are neglected, so the motor does not work well.
 As $K$ increases, $k^i_j$'s also increase, and when $K \approx$9.5 for
 $pK_a$=4.5 and $K \approx$10.5 for $pK_a$=5.5, the largest rate 
 $k_{\rm out}^{\rm L}$
 becomes comparable to $(\gamma \tau)^2/D$. A proton in the 
 left channel can dissociate before diffusion and state R is created.
 Thus $\bar {\Pi}_{\rm R}$ increases and is the largest among
 $\bar {\Pi}_i$'s in this region of $K$.

Next, for $pK_a$=4.5, $k_{\rm out}^{\rm R}$ becomes comparable to
 $(\gamma \tau)^2/D$ when $K \approx$12. 
Near this region of $K$, a proton in the right channel in state R can
 dissociate before diffusion, so state R becomes state E.  
Thus $\bar {\Pi}_R$ decreases
 until $\bar {\Pi}_E$ becomes larger than $\bar {\Pi}_R$.
As a result,
 the increase of $N({\rm H}^+)$ $(=\bar{\Pi}_R(\delta))$ 
is suppressed in this region,
 but $\langle v \rangle$ does not
 because the increase of $\bar {\Pi}_E$ cancels out the decrease of
 $\bar {\Pi}_R$ (See Figs.\ref{rot} and \ref{nh}). 
On the other hand, for $pK_a$=5.5, $\bar {\Pi}_R$ 
 continues to increase since
 $k_{\rm in}^{\rm R}$ is larger than $k_{\rm out}^{\rm R}$. Thus
 $N(H^+)$ and $\langle v \rangle$ increase monotonically. 

At last, when $K\gtrsim$16, all the transition rates become greater
 enough than $(\gamma \tau)^2/D$, so the matrix $\bf {\hat M}$ is reduced to
\begin{eqnarray}
&&\hspace{20ex} \bf{\hat M} \rightarrow
\nonumber\\
&& \left[
\begin{array}{cccc}
k_{\rm in}^{\rm R}k_{\rm out}^{\rm L}\gamma\tau
&0&0&-k_{\rm in}^{\rm R}D\eta_{\rm L}\\
k_{\rm out}^{\rm R}k_{\rm in}^{\rm L}\gamma\tau&0
&-k_{\rm out}^{\rm R}D\eta_{\rm L}e^{\eta_{\rm L}\delta}&0\\
k_{\rm out}^{\rm R}k_{\rm out}^{\rm L}\gamma\tau
&0&0&-k_{\rm out}^{\rm R}D\eta_{\rm L}\\
k_{\rm out}^{\rm R}k_{\rm out}^{\rm L}\gamma\tau&0
&k_{\rm out}^{\rm R}D\eta_{\rm L}e^{\eta_{\rm L}\delta}&0\\
k_{\rm in}^{\rm R}k_{\rm out}^{\rm L}-k_{\rm out}^{\rm R}k_{\rm in}^{\rm L}
&k_{\rm in}^{\rm R}k_{\rm out}^{\rm L}e^{\xi\delta}-k_{\rm out}^{\rm R}
k_{\rm in}^{\rm L}
&-k_{\rm in}^{\rm R}e^{\eta_{\rm L}\delta}
&-k_{\rm out}^{\rm R}\\
0&k_{\rm in}^{\rm R}k_{\rm in}^{\rm L}(1-e^{\xi\delta})
&-k_{\rm in}^{\rm R}e^{\eta_{\rm L} \delta}&
k_{\rm in}^{\rm R}\\
0&\xi k_{\rm in}^{\rm R}k_{\rm in}^{\rm L}(1-e^{\xi\delta})
&-k_{\rm in}^{\rm R}\eta_{\rm L}e^{\eta_{\rm L}\delta}&
-k_{\rm in}^{\rm R}\eta_{\rm L}\\
\delta&(e^{\xi\delta}-1)/\xi&0&0\\
\end{array}
\right.
\nonumber\\
&&\left.
\begin{array}{cccc}
0&k_{\rm out}^{\rm L}D\eta_{\rm R}&0&-D\eta_{\rm LR}\\
k_{\rm in}^{\rm L}D\eta_{\rm R}e^{\eta_{\rm R}\delta}&0
&-D\eta_{\rm LR}e^{\eta_{\rm LR}\delta}&0\\
0&-k_{\rm out}^{\rm L}D\eta_{\rm R}&0&D\eta_{\rm LR}\\
k_{\rm out}^{\rm L}D\eta_{\rm R}e^{\eta_{\rm R}\delta}&0
&D\eta_{\rm LR}e^{\eta_{\rm LR}\delta}&0\\
k_{\rm out}^{\rm L}e^{\eta_{\rm R}\delta}&k_{\rm in}^{\rm L}
&-e^{\eta_{\rm LR}\delta}&1\\
-k_{\rm in}^{\rm L}e^{\eta_{\rm R}\delta}&k_{\rm in}^{\rm L}
&-e^{\eta_{\rm LR}\delta}&1\\
-k_{\rm in}^{\rm L}\eta_{\rm R}e^{\eta_{\rm R}\delta}&
-k_{\rm in}^{\rm L}\eta_{\rm R}
&-\eta_{\rm LR}^+e^{\eta_{\rm LR}\delta})&-\eta_{\rm LR}\\
0&0&0&0\\
\end{array}
\right],
\end{eqnarray}
where 
\begin{eqnarray}
\eta_{\rm L}&=&\sqrt{(k^{\rm L}_{\rm in}+k^{\rm L}_{\rm out})/D},
\\
\eta_{\rm R}&=&\sqrt{(k^{\rm R}_{\rm in}+k^{\rm R}_{\rm out})/D},
\end{eqnarray}
and 
\begin{eqnarray}
\eta_{\rm LR}=\sqrt{(k^{\rm L}_{\rm in}+k^{\rm L}_{\rm out}
+k^{\rm R}_{\rm in}+k^{\rm R}_{\rm out})/D}.
\end{eqnarray}
(In this limit, $\eta_i^{\pm} \rightarrow \pm\eta_i$.)
From this, it is easily found that $C_{1-2}=O(10^{-2K})$, $C_{3-6}=
O(10^{-K})$, and $C_{7-8}=O(1)$, 
so $\langle v \rangle$ converges when $K$ diverges. 
The convergent value of $\langle v \rangle_{K \rightarrow \infty}$ is
given by 
\begin{eqnarray}
&&\langle v \rangle_{K \rightarrow \infty}  
\nonumber\\
&&= \frac
{\gamma^2 \tau^2 \delta [-bc+a\{e^{\gamma \tau \delta /D}
+b(e^{\gamma \tau \delta /D}-1)\}]}
{bc\{D(e^{\gamma \tau \delta /D}-1)-\gamma \tau \delta\}
+a[D-De^{\gamma \tau \delta /D}+\gamma \tau \delta 
\{e^{\gamma \tau \delta /D}+b(e^{\gamma \tau \delta /D}-1)\}]}, 
\nonumber\\
\end{eqnarray}
where $a=k^{\rm R}_{\rm in}/k^{\rm L}_{\rm out}$,
 $b=k^{\rm L}_{\rm in}/k^{\rm L}_{\rm out}$,
 $c=k^{\rm R}_{\rm out}/k^{\rm L}_{\rm out}$. 
 Comparing (\ref{nhp}) with (\ref{v}), it is also found that
$N({\rm H}^+)=O(10^{K/2})$. 
The divergent behavior of $N({\rm H}^+)$ is consistent with Fig. \ref{nh}.
 
\begin{figure}
\begin{center}
\includegraphics[width=7cm]{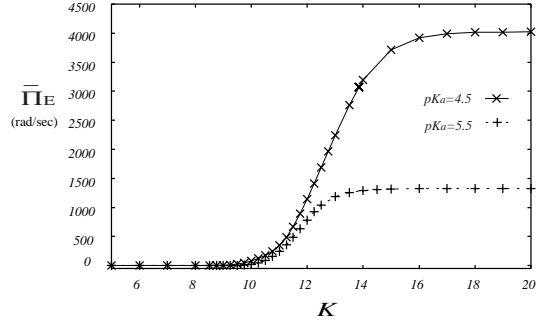}
\caption{The integrated flow of state E.
 The solid line with $\times$'s is the plot for $pK_a$=4.5 
and the dotted line with +'s is for $pK_a$=5.5. It increases monotonically
 both for $pK_a$=4.5 and 5.5. \label{ste}}
\end{center}
\end{figure}

\begin{figure}
\begin{center}
\includegraphics[width=7cm]{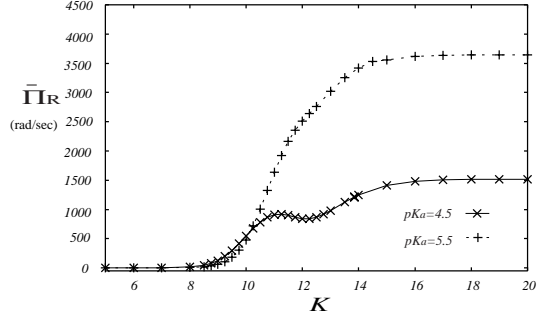}
\caption{The integrated flow of state R.
 The solid line with $\times$'s is the plot for $pK_a$=4.5
 and the dotted line with +'s
 is for $pK_a$=5.5. For $pK_a$=5.5, it increases
 monotonically. But for $pK_a$=4.5, there is a region where it 
 decreases. \label{str}}
\end{center}
\end{figure}

\begin{figure}
\begin{center}
\includegraphics[width=7cm]{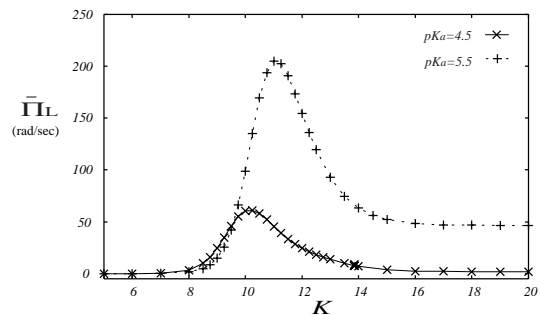}
\caption{The integrated flow of state L.
 The solid line with $\times$'s is the plot for $pK_a$=4.5
 and the dotted line with +'s is for $pK_a$=5.5. It has a peak
 for $K\approx$11 for $pK_a$=5.5 and $K\approx$10 for $pK_a$=4.5.
\label{stl}}
\end{center}
\end{figure}

We show the integrated flows of probability for state E and R, $\bar
{\Pi}_E$ and $\bar {\Pi}_R$, in Figs \ref{ste} and \ref{str}. 
Their behaviors are consistent with the above explanations.
We also see that ${\bar \Pi_{\rm F}}$ and ${\bar \Pi_{\rm L}}$ are much
smaller than ${\bar \Pi_{\rm E}}$ and ${\bar \Pi_{\rm R}}$ under the above
two sets of inequalities ($k^{\rm L}_{\rm out}>k^{\rm R}_{\rm in}
>k^{\rm R}_{\rm out}>k^{\rm L}_{\rm in}$ and 
$k^{\rm L}_{\rm out}>k^{\rm R}_{\rm out}
>k^{\rm R}_{\rm in}>k^{\rm L}_{\rm in}$).
(See Fig.\ref{stl}.)
Thus the qualitative behavior of $\langle v \rangle$ and $N(H^+)$ is
determined by only the states E and R.
This is also consistent with the above explanation which neglects
the state L and F.  


\section{Summary and Discussion}

We have investigated the rotatory molecular motor using the simply biased
diffusion model.
The model depends on four chemical reaction rates, $k_{\rm L}^{\rm in}$,
$k_{\rm L}^{\rm out}$, $k_{\rm R}^{\rm in}$, $k_{\rm R}^{\rm out}$,
the diffusion and the friction constants of the motor, $D$, $\gamma$,
and the load torque, $\tau$. 
We have solved the model analytically, and examined the relation between
these chemical reaction rates and the physical quantities, such as the
rotational velocity, the proton translocation rate, and the efficiency
of the energy transduction.
It is found that below such value $K$ that $k^{\rm L}_{\rm out}$, the largest
among $k^i_j$'s, becomes comparable with $(\gamma\tau)^2/D$, both
$\langle v \rangle$ and $N({\rm H}^+)$ are very low so this model does
not work well and result in a bad efficiency and that there exists an 
optimal value for the chemical reaction to maximize the efficiency of 
the energy transduction.
We also found that when the inequalities 
$k^{\rm L}_{\rm out}>(\gamma\tau)^2/D>k^{\rm R}_{\rm out}
>k^{\rm R}_{\rm in}>k^{\rm L}_{\rm in}$ hold, 
the increase of the proton translocation can become very
slow and the efficiency of the motor is enhanced.
This efficiency enhanced mechanism is naturally explained from this set of
inequalities. 

This model is based on the diffusion process like the thermal
ratchet model for linear molecular motors\cite{JAP}, so that the
efficiency enhance mechanism above may give a clue to examine whether diffusion
process dominates the motion of molecular motor essentially or not. 
The inequality is controlled by the dissociation constant of the proton
binding sites, $pK_a$, the proton concentration of both side of
the membrane, the proton diffusion constant, {\it etc}.   

The efficiency of this
model is still not high comparing with that reported in
Ref.\cite{NYK}.  
There are some possibilities which can explain this: 1) The definition
of efficiency may be still vague and controversial. 
In Ref.\cite{NYK}, the authors adapted a different definition of the efficiency
from ours.
Our definition coincides with that in
Ref.\cite{Sekimoto}, where the energetics of the Langevin equations and
the Fokker-Planck equations are carefully analyzed. 
2) There may be a necessity to reconsider the several relations between
physical quantities which we assume to be appropriate near equilibrium,
such as the Einstein's relation, since molecular motor is a
non-equilibrium system far from equilibrium. 
3) There may be another mechanism which enhance the efficiency. 
For instance, an electrostatic interaction between residues is discussed in
Ref.\cite{EWO}. 
Furthermore, experimentally conformational change of the motor protein
during motion is reported\cite{RG}. 
It is important to construct a model which gets along with this
experimental result. 
These may be the key to account for the high efficiency of energy
transduction of molecular motors.

\end{document}